\documentclass[12pt,twoside]{article}   
\usepackage[super,sort,comma]{natbib}
\usepackage{amsmath}
\usepackage{ragged2e}  

\usepackage{fancyhdr}		

\usepackage{showkeys}		

\usepackage[section]{placeins}   %

\usepackage{graphicx}

\makeatletter \renewcommand\@biblabel[1]{$^{#1}$} \makeatother
\newcommand{\cen}[1]{\begin{center} #1 \end{center}}

\usepackage{hyperref}
\hypersetup{ colorlinks,
    citecolor=blue,
    filecolor=blue,
    linkcolor=blue,
    urlcolor=blue
}

\usepackage{xcolor}

\definecolor{gray}{rgb}{0.6,0.6,0.6}
\definecolor{red}{rgb}{0.85,0,0}
\definecolor{green}{rgb}{0,0.85,0}
\definecolor{blue}{rgb}{0,0,0.85}
\definecolor{beige}{rgb}{0.92,0.87,0.78}
\usepackage[all]{hypcap}    

\begin{document}

\textcolor{red}{\bf THIS CONTENT IS PROTECTED BY A PATENT}

\cen{\sf {\Large {\bfseries A novel Boltzmann equation solver 
for calculation of dose  and fluence spectra distributions for 
proton beam therapy } \\  
\vspace*{10mm}
Oleg N Vassiliev, Radhe Mohan} \\
Department of Radiation Physics, The University of Texas
MD Anderson Cancer Center, Houston, TX
\vspace{5mm}\\
Version typeset \today\\
}

\pagenumbering{roman}
\setcounter{page}{1}
\pagestyle{plain}
Author to whom correspondence should be addressed.

 email: onvassiliev@mdanderson.org \\

\begin{abstract}
\noindent {\bf Background:}
The claim that Monte Carlo is the most accurate method is
a case of misattributed credit. This claim is based on experience
with advanced systems MCNPX,  Geant4 and EGS. 
These systems achieve remarkable performance because they
use most accurate physics, not because they use random numbers. 
The latter simplifies algorithms, but  contaminates the solution 
with random noise. Currently prevalent 
fast Monte Carlo algorithms retain this worst part while achieving
high computing speed  at the expense of the best part -
accurate physics. We employ an opposite strategy. We develop a
Boltzmann solver for protons that retains unchanged the physics of 
most  advanced Monte Carlo systems.  We eliminate random noise, 
because our solution method is deterministic.  Our method is also
applicable to heavier ions, helium and carbon, for example.\\  
 {\bf Purpose:} 
To develop a fast and accurate deterministic Boltzmann 
solver for  protons. It calculates dose distributions and fluence 
spectra. The spectra are needed for biological modelling.  
The main application is treatment planning of proton beam therapy.
\\
{\bf Methods:}
We solve the Boltzmann transport equation using an iterative procedure. 
Our algorithm accounts for Coulomb scattering and nuclear reactions. 
It uses the same physical models, as do the most rigorous Monte Carlo
systems. Thereby it achieves the same low level of systematic errors.
 Our solver does not involve random sampling. The solution is not
contaminated by statistical noise. This means that the overall 
uncertainties of our solver are lower than those realistically 
achievable with Monte Carlo. Furthermore, our solver is orders of 
magnitude faster. Its another advantage is that it calculates fluence 
spectra. They are needed for calculation of relative biological 
effectiveness, especially when advanced radiobiological models are used
that may present a challenge for other algorithms. \\
{\bf Results:}
We have developed a novel Boltzmann equation solver, have written 
prototype software, and completed its testing for calculations
in water. For 40-220 MeV protons we calculated fluence spectra, depth 
doses, three-dimensional dose distributions for narrow Gaussian beams. 
The CPU time was 5-11 ms for depth doses and fluence spectra at 
multiple depths. Gaussian beam calculations took 31-78 ms. All the 
calculations were run on a single Intel i7 2.9 GHz CPU. Comparison of 
our solver with Geant4 showed good agreement for all energies and 
depths. For the 1\%/1 mm $\gamma$-test the pass rate was 0.95-0.99.
In this test, 1\% was the difference between our and Geant4 doses at 
the same point. The test included low dose regions down to 1\% of the 
maximum dose.\\
{\bf Conclusions:}
Results of the study provide a foundation for achieving a high 
computing speed with uncompromised accuracy in proton treatment 
planning systems.  \\

\end{abstract}

\newpage     

\tableofcontents

\newpage

\setlength{\baselineskip}{0.7cm}      

\pagenumbering{arabic}
\setcounter{page}{1}
\pagestyle{fancy}
\section{Introduction}

The development of dose calculation algorithms for proton beam
therapy is largely limited to Monte Carlo based approaches.
Monte Carlo simulations can formally be considered as a method for 
solving the Boltzmann transport equation. This equation can also be 
solved with deterministic methods. The development of Acuros algorithm 
for x-rays  \cite{Va10}  and the results of its extensive 
testing that followed, prove that the deterministic approach is a viable
alternative to Monte Carlo. Although hadron physics is very different, 
potential advantages of the deterministic approach in treatment planning
software for protons are worth investigating. Also, to support research 
on biological optimizations of proton treatments and development for 
that purpose of advanced radiobiological models, reliable methods for 
calculation of additional physical characteristics, such as fluence 
spectra, are needed. This task deserves more attention than it is 
currently receiving. Our study addresses these matters. 

Early research on analytical methods for proton dose calculations 
\cite{Be93,De98} was based on Moli\'ere's theory of  
multiple scattering. This theory accurately predicts angular and radial 
dose distributions resulting from multiple Coulomb scattering (for 
example\cite{Be93}). However, for radiotherapy dose calculations
energy straggling and nuclear reactions must also be accounted for.
To accomplish this, the method by Deasy\cite{De98}  requires 
depth doses calculated with Monte Carlo.

Regarding algorithms designed for proton therapy treatment 
planning, pencil beam (PB) have been the most common algorithm type. 
Poor accuracy of such algorithms is well documented\cite{Gr14,Sc15} 
Taylor et al.\cite{Ta17} reported results of 
a multi-institutional study, in which a lung phantom  was irradiated and
doses in the target volume were measured. Two commercial PB algorithms 
overestimated the dose to the center of the target volume by 7.2\% on 
average. Elsewhere in the target volume, PB calculations overpredicted 
the dose by up to 46\%. 

Full Monte Carlo algorithms, such as those used in MCNPX\cite{Pe11}
and Geant4\cite{Ag03,Al16} , are highly   
accurate and reliable. However, they are too slow for routine treatment 
planning. Several algorithms described as "fast Monte Carlo" have been 
proposed. Review of every such algorithm is not feasible. We will 
discuss only algorithms implemented in commercial treatment planning 
software. They represent some of the best algorithms of this type. We 
will consider Acuros PT by Varian Medical Systems (Palo Alto, CA) and an 
algorithm by RaySearch Laboratories (Stockholm, Sweden). 

In Acuros PT, the level of statistical uncertainties in the target 
volume that can be achieved within a few minutes of calculation time is 
about 2\%\cite{Li17}. Statistical uncertainties in organs at risk
 are even higher, because they receive lower doses. The increase is by a 
factor of dose$^{-1/2}$. Systematic errors are not reported. They are 
difficult to quantify. They 
further increase the overall error. Systematic errors 
arise from approximations made in physical models and algorithms. The 
Acuros PT algorithm is based on the Fokker-Planck equation 
\cite{Li17,Va21}. It is a small-angle form
 of the full Boltzmann equation. In this approximation the energy loss 
distributions is Gaussian\cite{Mo81,Va17}. However, the 
Gaussian distribution of energy losses was not implemented. Instead, a
new algorithm was introduced. It is inconsistent with the Fokker-Planck
 equation, and required additional approximations. The Fokker-Planck 
approximation is not used in the standard Monte Carlo systems, such as 
MCNPX (recent versions) and Geant4. They rely on more rigorous physics. 
This reflects the prevailing view that the Fokker-Planck approximation 
is incompatible with the objective of achieving high accuracy that 
motivates the choice of Monte Carlo over other methods. 

In Acuros PT, energy deposited in a voxel is calculated as a product 
of the stopping power and path length of the particle in the voxel. This 
method introduces an error, because the stopping power changes as the 
particle traverses the voxel. Errors associated with these 
approximations are small. However, a simple method that eliminates this
error exists. It is based on lookup of inverse range versus energy 
table. Acuros PT does not calculate fluence spectra that are needed for 
calculation of RBE. Nor does it calculate LET spectra or average LETs 
used in more basic radiobiological models. 

Lin etal\cite{Li17} compared Acuros PT dose calculations with measurements 
and Monte Carlo simulations performed using TOPAS software 
\cite{Pe12}. The measurements were done in water. Discrepancies 
between Acuros PT and the measurements reached 4\% for the field size 
factor, 16\% for the penumbra width, and 15\% for spot sizes of pencil 
beams. Experimental uncertainties did contribute to these discrepancies. 
However, this was not the main factor, because TOPAS achieved 
substantially better agreement with experiment than did Acuros PT. This
difference in performance between the two Monte Carlo codes shows the 
advantage of more rigorous physical models implemented in TOPAS, which 
is based on Geant4.

In the proton Monte Carlo algorithm by RaySearch Laboratories 
\cite{Ra22}, for each beam the mean standard deviation
over all voxels having a dose above 50\% of the maximum dose is output. 
Statistical uncertainty of 0.5\% can be achieved within half a minute of
calculation time\cite{Sc19B}. The mean value is a poor 
measure, because it does not exclude the possibility of much higher 
errors in some voxels. For safety reasons, the maximum error should be
reported instead. Most normal organs and tissues receive less than 50\% 
of the maximum dose. Hence, dose uncertainties are unknown. This brings 
the question: how anyone can approve a treatment plan when the 
calculated dose is close to the tolerance level, and the dose confidence
interval is not known? This situation is not uncommon.

 Angular 
distributions of protons are modelled using the theory of Goudsmit and 
Saunderson with a screened Rutherford cross section. The latter was 
designed in previous studies for electrons. When applied to protons 
this formalism required a 7\% correction, uniformly applied to all 
scattering angles, independent of proton energy or medium. This cannot 
be accurate. Protons are much heavier than electrons and have an 
opposite charge. Energy loss straggling is modelled in the Bohr 
approximation. The formula for $\sigma^2$ that it provides is 
inaccurate , because it is based on an outdated scattering model. 
Going this far back  in time cannot be justified when much improved 
models are available. This is worrse than the Fokker-Planck model. 
The latter produces an accurate $\sigma^2$, when properly 
implemented. Energy and angular distributions are modelled as 
described above only until a proton slows down to an energy that 
corresponds to a residual range in water of 9 mm. "Below this energy the
proton is transported without considering multiple scattering and energy
loss straggling."\cite{Ra22}. However, near the end 
of a proton track, dose, dose gradient, LET and RBE reach their maxima, 
and lateral scattering remains strong. The last 9 mm of a proton track
is not a location where lowering the accuracy to save computing time is
 justifiable. Energy deposited in a voxel is calculated without 
 accounting for  straggling or multiple scattering \cite{Ru20} .
The above "acceleration techniques" exemplify the physics oversimplification 
approach that is the basis of the fast Monte Carlo trend. In addition 
to dose distributions, the algorithm calculates distributions of the 
dose-average LET. It is used in some basic models to calculate RBE. More
 advanced RBE models require more detailed information about beam 
physics. 

Performance testing of the RaySearch Monte Carlo software 
\cite{Sa17} reported ±3\% agreement with experiment for depth 
dose in water. In a $\gamma$-index test (3\%,3 mm), in a heterogeneous 
slab phantom, the pass rate was 90\% for six out of seven planes. 
Schreuder et al.\cite{Sc19A} reported somewhat better results of 3\%/3 mm 
$\gamma$-tests for a neck phantom and a water-filled breast phantom. 
Still, for the neck phantom in 2 out of 8 tests the pass rate was below 
89\%. For the 2\%/2 mm criterion the average pass rate for the neck 
phantom was 0.819. In a similar study with a lung phantom 
Schreuder et al.\cite{Sc19B} reported the pass rate of 92\% (3\%/3 mm) for a 
single beam irradiation. For the same setup, RaySearch Monte Carlo 
software underestimated dose in the distal region of the target volume 
by 18\%. Ruangchan et al.\cite{Ru020} compared proton dose calculations with
measurements in a heterogeneous phantom comprised of bone and lung or 
soft tissue blocks. Differences in mean target doses were within 
$\pm3 \%$. Outside the target the maximum dose difference was 9\%. 

These fast Monte Carlo algorithms offer better performance than do PB 
algorithms. They, however, do not consistently comply with the 
AAPM Task Group 185 Report\cite{Fa20} that recommends achieving
the pass rate of 95\% for the 3\%/3 mm $\gamma$-test. Furthermore, such 
algorithms do not eliminate the risk of dose miscalculation by up to
$\sim$10\%. They are a long way from reaching the 2\%/2 mm level of 
accuracy that the AAPM Task Group 157\cite{Ma20} recommends for 
Monte Carlo dose calculation for photon and electron beams. Of course, 
there are variations in the performance between different algorithms in 
the "fast Monte Carlo" category. However, extensive statistical sampling 
that is required to control uncertainties throughout the irradiated 
volume limits the computational efficiency achievable with Monte Carlo. 
To advance beyond this efficiency limit, it is worth exploring 
alternative computational approaches. 

In recent years progress has been made in applying deep 
leaning (DL) techniques to proton dose calculations
\cite{(No20,Ja21,Wu21,Pa22}. Algorithms of
this type can achieve a high calculation speed. The downside is that
commissioning of software for the clinic requires an extraordinary 
amount of work. This may include, for example, dose calculations for 
80000 different geometries \cite{Pa22}. It is not feasible to
complete this amount of calculations within a reasonable time using a 
full Monte Carlo algorithm. This limits the accuracy of DL algorithms to 
that of fast Monte Carlo. The DL models are affected by statistical 
uncertainties in the Monte Carlo data. Therefore, they are prone to 
large errors in low dose regions. DL algorithms are not concerned at all
with direct modelling of real physical processes. Hence, they are less
reliable than physics-based algorithms. This means that they require 
more extensive testing to reasonably eliminate the risk of a clinically 
significant dose miscalculation. Finally, existing DL algorithms do not
offer such data as fluence spectra.

Bedford\cite{Be23} developed an algorithm that is described as a solution of 
the Boltzmann transport equation by the discrete ordinates method. 
The study lacks proper validation or performance data. For comparison, 
please see the multotude of papers on evaluation of Acuros XB algorithm.
It is a stretch to call this proton algorithm a discrete ordinates 
solution of the Boltzmann equation. The main part of the solution, 
unscattered fluence, is approximated by a formula of the type used in 
pencil beam algorithms.This formula is not a solution of the Boltzmann 
equation. Unscattered fluence is the source of scattered protons. 
Therefore, a more accurate method used to find scattered fluence hardly 
improves the overall accuracy. 

A different approach was developed in Burlacu et al.\cite{Bu23,Bu24}. 
Neither 
study describes a Boltzmann solver. The Boltzmann equation can be written in 
a number of differebnt forms. However, as a minimum, it must include the 
collision integral. In both papers, the collision integral is approximated by partial 
derivatives. This puts these algorithms in the same category as Fokker-Planck 
solvers. Both studies solve equations for fluence, but neither reports validation
 data for fluence spectra. This is an importantand sensitive test. 
The Fokker-Planck model fails it. It does not produce asymmetric fluence
 spectra of the shape shown ib Figs. 3-6. Nuclear reactions are not 
modelled. Without nuclear reactions, high accuracy is not achievable 
\cite{Pa02}.

In the present study we introduce a novel solver of an appropriate form
of the Boltzmann transport equation, a deterministic Boltzmann solver 
(DBS). The solver is intended for treatment planning of proton beam 
therapy. Our implementation of the solver uses the same physical models
as does the most rigorous Monte Carlo software. Thereby it achieves the 
same low level of systematic uncertainties. Our DBS does not involve
random sampling. This eliminates statistical uncertainties. Hence, its
overall uncertianties are lower than those of the best Monte Carlo 
software. For the same reason, our solver is orders of magnitude faster.
Another advantage of our DBS is that it also calculates fluence spectra.
They are intended for RBE calculations using either LET-based models or 
more sophisticated models. We report substantial data on the performance 
of our DBS for calculations in water. We compare our calculation results 
with Monte Carlo simulations performed with Geant4 software 
\cite{Ag03, Al16}.

\section{Methods } 
\subsection{Foundations of the method} 

Our new solver is based on the Lagrangian form of the Boltzmann 
transport equation\cite{(Va17}:

\begin{equation*}
\frac{\partial}{\partial t}\Phi\left(\vec{r},\vec{\Omega},E,t\right)+ 
\sigma\left(\vec{r},E\right)\Phi\left(\vec{r},\vec{\Omega},E,t\right)=
\end{equation*}
\begin{equation}
\int_{0}^{\infty}\mathrm{d}E^\prime \int_{4\pi}
\mathrm{d}\vec{\Omega}^\prime
\sigma_s\left(\vec{r}\:;\:\vec{\Omega}^\prime, E^\prime \to
\vec{\Omega}, E\right)
\Phi\left(\vec{r},\vec{\Omega}^\prime,E^\prime,t\right).
\end{equation}

\noindent Notations: $t$, path length; $\Phi$, fluence; $\vec{r}, 
\vec{\Omega}, E$, phase coordinates of the particle: its location, 
direction of travel (a unit vector), kinetic energy; $\sigma$, total 
interaction cross section; $\sigma_s$, double differential scattering 
cross section; proton scattering has azimuthal symmetry, this means that
$\sigma_s$ is not a function of two vectors $\vec{\Omega}$ and 
$\vec{\Omega}^\prime$, but a function of the cosine of the scattering 
angle, $\cos{\theta}=\vec{\Omega}\cdot\vec{\Omega}^\prime$.

This form of the Boltzmann equation, although uncommon, is not new, it
was previously discussed in the literature, for example by Wienke 
\cite{Wi74,Wi82}. The well-known algorithm for x-rays, 
Acuros\textsuperscript{\textregistered}\cite{Va10}, 
also solves the Boltzmann equation, but in a different form that 
can be classified as Eulerian. Developers of Monte Carlo algorithms 
usually simply imitate the actual physical processes. However, Monte 
Carlo algorithms for particle transport can be formally derived as 
solvers of the integral form of the Boltzmann equation 
\cite{Va17}.

Equation (1) is integro-differential. To solve it we use an iterative 
procedure based on a line integration method. We integrate Eq. (1) 
along the particle path step-by-step, starting at $t=0$ and making 
finite steps $\Delta t$ until all particles stop. The step size 
$\Delta t$ is not random. It is variable and optimized to achieve the 
best balance between computing speed and accuracy. The physics of proton
interactions with matter is such that the difference between the proton 
path length $t$ and the corresponding penetration depth in matter $z$ is 
small. This difference is characterized by the detour factor defined as 
the ratio of projected range (i.e. average value of the depth to which a 
charged particle penetrates in the course of slowing down to rest) and 
continuous slowing down range (the total path length). For protons with 
energies from 10 MeV to 250 MeV propagating in water the detour factor 
is 0.9980-0.9989 \cite{NI22}. That is, the difference between $t$ and 
$z$ is only 0.1-0.2\%. Hence, we assume $t=z$. 
This approximation is applicable to other materials. For 
example, for protons with 1 mm range $z/t=0.998$ for ICRU compact 
boone, and $z/t=0.981$  for tungsten. The 2\% difference for tungsten
can be corrected for by pretabulating $ t(z)$ function.
This approximation does not mean that 
we neglect lateral scattering of protons or recoil particles. Our 
solver calculates angular and radial distributions of proton fluence. Both 
distributions are very narrow. The approximation $t=z$ is even more 
accurate for heavier particles used in radiotherapy, such as helium and 
carbon ions. Therefore our methods can be extended to include such 
particles. 

\subsection{Multiple Coulomb scattering} 

\subsubsection{Fluence spectra}
\hfill\\

\noindent
\textbf{\textit{Iterative procedure.}} We use an iterative procedure to 
calculate fluence spectra:

\begin{equation}
\Phi_{i+1}(E)=\int_E^\infty \Phi_i(E^\prime)\Phi_i(E \vert E^\prime)
              \mathrm{d}E^\prime.
\end{equation}

\noindent This equation is the total expectation formula. Here 
$\Phi_{i+1}(E)$ and $\Phi_i(E)$ are fluence spectra at depths $z_{i+1}$ 
and $z_i$, $z_{i+1}>z_i$; $\Phi_i(E \vert E^\prime)$ is the conditional 
expectation, i.e. it is the fluence spectrum at depth $z_{i+1}$ produced 
by protons that had energy $E^\prime$ when they were at depth $z_i$. If 
the range of a proton with energy $E^\prime$ is less than the step size 
$\Delta t_i=z_{i+1}-z_i$, then $\Phi(E \vert E^\prime)=0$. 

To calculate $\Phi_i(E\vert E^\prime)$, we use Vavilov distribution 
\cite{Va57,Va17}. Our algorithm allows to use Vavilov 
distribution for all steps, or only for a few first steps. Because 
calculation of Vavilov distribution is relatively slow, for optimal 
performance we recommend using it only for the first spectrum 
$\Phi_1(E\vert E^\prime)$. Vavilov distribution has been previously used 
in advanced Monte Carlo software \cite{Pe11}. Here we introduce a 
method for using it in a deterministic solver.

For all the steps that do not use Vavilov distribution, we approximate
the conditional distribution $\Phi_i(E \vert E^\prime)$ with a normal
distribution. This approximation is based on the asymptotic properties 
of Vavilov distribution that tends to a normal distribution as the step 
size $\Delta t_i$ increases \cite{Va57, Va17}. 
Hence, the step size $\Delta t_i$ should be 
sufficiently large for the normal distribution to be an accurate 
approximation of the exact Vavilov formula. On the other hand, 
calculations are substantially simplified when particle energy loss per
step is small, which is achieved by limiting the step size from above. 
Balancing these two conflicting requirements determines the optimal step 
size. We determined optimal step sizes by performing our 
calculations with various step sizes and comparing the results with 
Monte Carlo simulations. Equation (1) does not account for loss of 
protons in nuclear reactions. We discuss this later.

\hfill\\
\noindent\textbf{\textit{Vavilov distribution.}} This is an overview 
only. For derivation of the distribution we refer to Vavilov\cite{Va57}
 and  Vassiliev\cite{Va17}. Vavilov distribution is a solution of the multiple 
scattering problem for charged particles that travel a distance $t$ such 
that the energy losses are much smaller than the initial energy of the 
particles. The solution is based on a relativistic form of the 
Rutherford formula for the scattering cross section
\cite{Va57, Se88}:

\begin{equation}
\sigma_s(E,\Delta E)=\frac{\xi(E)}{(\Delta E)^2}
  \left[1-\beta^2\frac{\Delta E}{(\Delta E)_{max}}\right],
\end{equation}

\begin{equation}
(\Delta E)_{max}=\frac{2mc^2 \beta^2}{1-\beta^2},
\end{equation}

\begin{equation}
\xi(E)=2\pi r_e^2 \rho N_A \frac{mc^2}{\beta^2}\frac{Z}{A}.
\end{equation}

\noindent Notations: $mc^2$ is the electron rest energy; $\beta=v/c$ is 
the ratio of proton velocity to the speed of light; $r_e$ is the 
classical electron radius; $\rho$ is the mass density of the material;
$N_A$ is the Avogadro's number; $Z$ is the number of electrons per 
molecule; $A$ is the molar mass of the material.

If protons start at $t=0$ with an initial energy $E_0$ and travel a 
distance $t$, then the fluence spectrum expressed in terms of energy 
lost $Q=E_0-E(t)$ is 

\begin{equation}
\Phi\left(t,\!Q\right)=
\frac{\exp{\left[k\left(1+\gamma v^2/c^2\right)\right]}}
{\pi \left(\Delta E\right)_{\mathrm{max}}}
\int_0^{\infty}
\mathrm{e}^{kf_1\left(y\right)}
\cos{\left[\lambda_1 y+k f_2\left(y\right) \right]}
\mathrm{d}y,
\end{equation}

\noindent where 

\begin{equation}
f_1\left(y\right)=\left(\frac{v}{c}\right)^2\left[\ln{y}-
\mathrm{Ci}\left(y\right) \right]-y\mathrm{Si}\left(y\right)-\cos{y},
\end{equation}

\begin{equation}
f_2\left(y\right)=y\left[\ln{y}-\mathrm{Ci}\left(y\right) \right]+
\left(\frac{v}{c}\right)^2\mathrm{Si}\left(y\right)+\sin{y},
\end{equation}

\begin{equation}
k=\frac{\xi(E)t}{(\Delta E)_{max}},
\end{equation}

\begin{equation}
\lambda_1=\frac{Q-\bar{Q}(E,t)}{(\Delta E)_{max}}-k
        \left[1+\left(\frac{v}{c}\right)^2-\gamma\right],
\end{equation}

\noindent $\gamma$ is Euler's constant and $\bar{Q}(E,t)$ is the
average energy lost over distance $t$ by a proton with energy $E$.
Calculation of one spectrum, $\Phi_i(E\vert E^\prime )$, using 
Eqs. (6)-(10) takes a fraction of a second. However, our algorithm 
avoids multiple calculations of Vavilov spectra to minimize the overall 
computing time. To this end, for those steps for which we chose not to 
use Vavilov distribution, we approximate $\Phi_i(E\vert E^\prime)$ 
with a normal distribution.

\hfill\\
\noindent\textbf{\textit{Normal distribution.}}
If we approximate $\Phi_i(E\vert E^\prime)$ with a normal distribution,
then we need to calculate its center $\bar{E}$ and width $\sigma^2$. We 
use the continuous slowing down approximation (CSDA) to calculate 
$\bar{E}$. First, we tabulate range versus proton energy, $R(E)$. If a 
proton starts with an energy $E^\prime$ and travels distance $\Delta t$,
then its average  energy $\bar{E}$ at the step end is\cite{Va17}:

$
 \bar{E}=
\begin{cases}
R^{-1}\left[R\left(E^\prime \right)-\Delta t \right] ; 
      & \Delta t < R\left(E^\prime\right)    \\
    0; & \Delta t \ge R\left(E^\prime\right),
\end{cases}
$

\noindent where $R^{-1}$ is the inverse function. The corresponding 
distribution width is\cite{Ro90} :

\begin{equation}
\sigma^2=\xi \Delta t (\Delta E)_{max}(1-\beta^2/2).
\end{equation}

\hfill\\
\noindent\textbf{\textit{Interpolation of fluence spectra.}}
Fluence spectra are calculated at depths $z_1,z_2, \ldots, z_n$ chosen 
so as to optimize the speed and accuracy of the calculations. To find
fluence spectra at any other depth, $z_1<z<z_n$, we use an interpolation 
method. When proton fluences at all the depths $z_i$ are calculated, 
for each depth the average proton energy $\bar{K}_i$ is calculated
using Eq. 11. By the same method the average proton energy $\bar{K}$ 
at depth $z$ is also calculated. Then, two depths $z_m$ and $z_{m+1}$ 
nearest to $z$ on both sides ($z_m<z<z_{m+1}$) are determined, and the 
weight and two energy shifts  are calculated as follows:

\begin{equation}
w=\frac{\bar{K}_m-\bar{K}}{\bar{K}_m-\bar{K}_{m+1}}.
\end{equation}

\begin{equation}
\Delta K_m=\bar{K}_m-\bar{K}.
\end{equation}

\begin{equation}
\Delta K_{m+1}=\bar{K}-\bar{K}_{m+1}.
\end{equation}

\noindent Then the interpolation formula is

\begin{equation}
\Phi(z,E)=w\Phi(z_m,E+\Delta K_m)+(1-w)\Phi(z_{m+1},E-\Delta K_{m+1}).
\end{equation}

\noindent This interpolation takes a fraction of a second and is quite
accurate, as Fig. 1 illustrates. 

\begin{center}
\includegraphics[width=110mm]{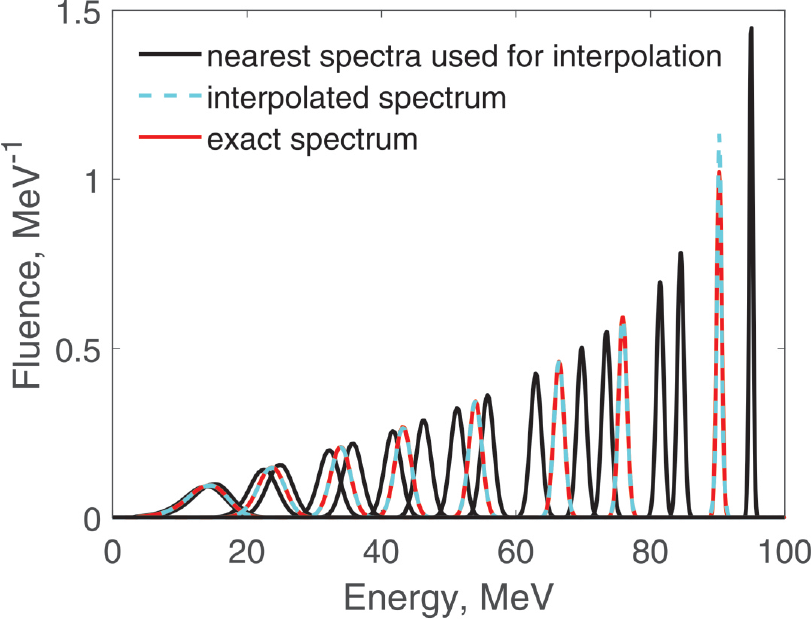}
\end{center}
\begin{flushleft}
    Figure 1. Calculation of fluence spectra at a given
    depth by interpolation between two precalculated spectra
    at nearby depths. The initial proton energy was 100 MeV.
\end{flushleft}
\vspace{0.2cm}


\subsubsection{Angular and radial distributions.}
\hfill\\

\noindent
\textbf{\textit{Iterative procedure for angular distributions.}} 
Angular distributions of fluence $\Phi(\vec{\Omega})$ are 
calculated  iteratively using the following formula:

\begin{equation}
\Phi_{i+1}(\vec{\Omega})=\int\Phi_i(\vec{\Omega}^\prime)
\Phi_i(\vec{\Omega}^\prime\cdot\vec{\Omega})
              \mathrm{d}\vec{\Omega}^\prime.
\end{equation}

\noindent Here $\Phi_{i+1}(\vec{\Omega})$,
$\Phi_i(\vec{\Omega}^\prime)$ are angular distributions of fluence at 
depths $z_{i+1}$ and $z_i$, $z_{i+1}>z_i$; 
$\Phi_i(\vec{\Omega}^\prime\cdot\vec{\Omega})$ is the angular 
distribution of fluence at depth $z_{i+1}$ produced by protons that were
travelling in direction $\vec{\Omega}^\prime$ when they were at depth 
$z_i$. We use Moli\`ere distribution (next paragraph) to calculate this 
distribution. To calculate the integral in Eq. (17), we note that

\begin{equation}
\vec{\Omega}=(\theta,\phi).
\end{equation}

\begin{equation}
\vec{\Omega}^\prime\cdot\vec{\Omega}=\cos{\Theta}=
\sin{\theta}\sin{\theta^\prime}\cos{(\phi-\phi^\prime)}
+\cos{\theta}\cos{\theta^\prime},
\end{equation}

\noindent where $\Theta$ is the angle between directions
$\vec{\Omega}$ and $\vec{\Omega}^\prime$. Because of the azimuthal
symmetry of the problem, we can set $\phi=0$. Then Eq. (17) can be 
written as follows:

\begin{equation}
\Theta(\theta,\theta^\prime,\phi^\prime)=
\cos^{-1}{(\sin{\theta}\sin{\theta^\prime}
\cos{\phi^\prime}+\cos{\theta}\cos{\theta^\prime)}}.
\end{equation}

\begin{equation}
\Phi(t+\Delta t,\theta)=\int_0^\infty
\Phi(t,\theta^\prime)\theta^\prime\mathrm{d}
\theta^\prime\int_0^{2\pi}\Phi(\Delta t,
\Theta(\theta,\theta^\prime,\phi^\prime))
\mathrm{d}\phi^\prime.
\end{equation}

\noindent In Eq. (21) we used a small angle approximation for 
$\theta^\prime$ \cite{Va17}. To calculate the integral over 
$\theta^\prime$ we use the Legendre quadrature. For the integral 
over $\phi^\prime$ we use the identity

\begin{equation}
\int_0^{2\pi}f(\cos{\phi})\mathrm{d}\phi=
2\int_{-1}^1 \frac{f(y)}{\sqrt{1-y^2}}\mathrm{d}y,
\end{equation} 
\noindent and then the Chebyshev quadrature.

\hfill\\
\noindent\textbf{\textit{Moli\`ere distribution for a small step.}}
To calculate angular distribution of protons for a small step $\Delta t$ 
we use Moli\`ere distribution. Here we give only a summary. Full 
details are given in Berger\cite{Be93} and Vassiliev\cite{Va17}. 
The angular  distribution is calculated as follows:

\begin{equation}
\Phi(\Delta t,\theta)=\frac{1}{2\pi\chi_c^2 B}\sum_{n=0}^{\infty}
               \frac{1}{B^n}f^{(n)}\left(\frac{\theta}
              {\chi_c\sqrt{B}} \right).
\end{equation}

\begin{equation}
f^{(n)}(\vartheta)=\frac{1}{n!}\int_0^\infty u\mathrm{d}u 
        J_0(\vartheta u)\exp{\left(-\frac{u^2}{4}\right)}
        \left[\frac{u^2}{4}\ln{\left(\frac{u^2}{4}\right)} \right]^n,
\end{equation}

\noindent where $\Delta t$ is the step size, $\theta$ is the 
scattering angle and $J_0$ is  the Bessel function of the first kind of 
order zero. Functions $f^{(n)}(\vartheta)$, where

\begin{equation}
\vartheta=\frac{\theta}{\chi_c \sqrt{B}},
\end{equation}

\noindent are pretabulated. Parameter $\chi_{c}^2$ for a chemical 
element $j$ is

\begin{equation}
    \chi_{c,j}^2=\frac{4}{3}\pi N_A \left(\frac{r_e}{M}\right)^2 
    \left[\frac{\tau+1}{\tau(\tau+2)}\right]^2 \Delta t,
\end{equation}

\noindent and for a compound it is
   \begin{equation}
   \chi_c^2=\sum_j w_j \chi_{c,j}^2.
   \end{equation} 

\noindent Notations: $w_j$ is the weight fraction of the $j$-th element;
$N_A$ is Avogadro's number; $M$ is the proton mass; $r_e$ is the 
classical electron radius; $\tau=E/Mc^2$. Parameter $B$ is found by 
solving this equation for a given $b$:
\begin{equation}
 b=B-\ln{B}.
\end{equation}

\noindent Parameter $b$ is calculated as follows:

\begin{equation}
 b=\ln{\frac{\chi_c^2}{\chi_a^2}}+1-2\gamma.
\end{equation}

\begin{equation}
\ln{\chi_a^2}=\frac{4\pi N_A}{\chi_c^2}\left(\frac{r_e}{M}\right)^2
\left[\frac{\tau+1}{\tau(\tau+2)}\right]^2
\sum_j w_j \chi_{a,j}^2 \frac{Z_j^2}{A_j}.
\end{equation}

\begin{equation}
\chi_{a,j}^2=\frac{\Delta t}{3}\left(\ln{G_{s,j}}-\frac{F_j}{Z_j} 
             \right).
\end{equation}

\begin{equation}
G_{s,j}=Z_j^{2/3}\left(\frac{\alpha}{M C_{TF}}\right)^2 
\left[1.13+3.76\left(\frac{\alpha Z_j}{\beta}\right)^2\right]
\frac{k_{HF,j}}{\tau(\tau+2)}.
\end{equation}

\noindent For calculations in water, for example:
\begin{equation}    
F_{j,hydrogen}=\ln{\left[1130\frac{\beta^2}{(1-\beta^2)^2}\right]}+
               3.6-\frac{\beta^2}{2}.
\end{equation}
\begin{equation}
F_{j,oxygen}=\ln{\left[1130\frac{8^{-4/3}\beta^2}
               {(1-\beta^2)^2}\right]}+
               5.8-\frac{\beta^2}{2}.  
\end{equation}

\noindent Notations: $Z_j$ is the atomic number of element $j$;
$\gamma$ is Euler's constant; $\alpha$ is the fine structure constant; 
$C_{TF}=$0.88534 is the Thomas-Fermi constant; $k_{HF,j}$ is the 
Hartee-Fock correction for element $j$\cite{Be93}.

\hfill\\
\noindent
\textbf{\textit{From angular distribution to radial distribution.}}
Radial distribution of fluence $\Phi(z,\rho)$ is derived from
angular distribution $\Phi(z,\vec{\Omega})$ using an adaptation for 
step-wise application of a technique introduced by Moli\`ere 
\cite{Mo55} and discussed by Berger\cite{Be93}. At depth $z_0=0$ we have 
$\bar{\rho}_0=\bar{\theta}_0=0$, where the bar indicates the average 
value. At depth $z_1>z_0$, we have 
$\bar{\rho}_1=\frac{1}{2} \tan{(\bar{\theta}_1)} (z_1-z_0)$. At depth
$z_{i+1}$ we have:

\begin{equation}
\bar{\rho}_{i+1}=\bar{\rho}_i+\frac{1}{2}
\left[\tan{(\bar{\theta}_{i+1})}+
\tan{(\bar{\theta}_i)}\right]
 (z_{i+1}-z_i).
\end{equation} 

\noindent The formula for conversion from angular to radial
distributions is

\begin{equation}
\Phi(z_i,\rho)=\Phi(z_i,{
\bar{\theta_i}}\rho/\bar{\rho}_i).
\end{equation}

\hfill\\
\noindent
\textbf{\textit{Interpolation of radial dose distributions.}}
If we need to find $\Phi(z,\rho)$ given $\Phi(z_i,\rho)$ and 
$\Phi(z_{i+1},\rho)$, where $z_i \le z \le z_{i+1}$, then the 
interpolation formula is:

\begin{equation}
w=\frac{z-z_i}{z_{i+1}-z_i}.
\end{equation}

\begin{equation}
\Phi(z,\rho)=(1-w)\Phi(z_i,\rho)+w\Phi(z_{i+1},\rho).
\end{equation}

Our solver calculates 
fluence distributions for the delta source. The result is,
in fact, Green's function of the Boltzmann equation. To find a solution
 for an arbitrary source, Geen's  function is multiplied by the source 
function and the product is integrated over the phase space. An 
example if such intgration is given in the next section.

\hfill\\
\noindent
\textbf{\textit{Narrow Gaussian beam. Radial distribution of fluence.}}
An incident proton fluence has a two-dimensional normal distribution:

\begin{equation}
\Phi_G(0,\rho)=\frac{1}{2\pi\sigma^2}
\exp{\left(-\frac{x^2+y^2}{2\sigma^2}\right)}=
\frac{1}{2\pi\sigma^2}
\exp{\left(-\frac{\rho^2}{2\sigma^2}\right)}.
\end{equation}

\noindent At depth $z$ the radial distribution of fluence is
\begin{equation}
\Phi_G(z,\rho)\!=\!\frac{1}{2\pi\sigma^2}
\!\int_{-\infty}^\infty \!\!\!\!\! \mathrm{d}x^\prime
\!\int_{-\infty}^\infty \!\!\!\!\! \mathrm{d}y^\prime
\exp{\left[-\frac{(x\!-\!x^\prime)^2\!+\!
(y\!-\!y^\prime)^2}
{2\sigma^2}\right]}\Phi(z,x^\prime,y^\prime).
\end{equation}

\noindent Using azimuthal symmetry of the problem, we set $y=0$, and 
then switch to polar coordinates

\begin{equation}
\Phi_G(z,\rho)\!=\!\frac{1}{2\pi\sigma^2}
\!\int_0^\infty \!\!\!\!\! \rho^\prime\mathrm{d}\rho^\prime
\Phi(z,\rho^\prime)
\!\int_0^{2\pi} \!\!\!\!\! \mathrm{d}\phi^\prime
\exp{\left[\frac{2\rho\rho^\prime\cos{\phi^\prime}
-\rho^2-(\rho^\prime)^2}
{2\sigma^2}\right]}.
\end{equation}

\noindent For the integral over $\rho^\prime$ we use the Legendre 
quadrature. We transform the integral over $\phi^\prime$ using Eq. (22), 
and then use the Chebyshev quadrature.

\subsection{Nuclear interactions} 

The equation for secondary protons is a separate e
quation, in which  primary fluence is the source. Hence,
nuclear processes are modelled after multiple Coulomb scattering 
calculations using Eq. (2) are completed. At that point we have proton 
fluence spectra $\Phi_i(E)$ calculated at all depths 
$z_1,z_2, \ldots, z_n$. We account for three nuclear processes: 
elastic interactions of protons with hydrogen atoms; elastic 
interactions of protons with atoms heavier than hydrogen (C,N,O, etc.); 
inelastic interactions of protons with atoms heavier than hydrogen. 
Inelastic interactions of protons with hydrogen atoms are negligible. 
The cross sections for these reactions are $\sigma_{el}^H(E)$, 
$\sigma_{el}^A(E)$, and $\sigma_{in}^A(E)$, respectively. The 
cross-sectional data were compiled from several sources
\cite{Ba89,Ar97, Ar98,Ge18}.

The total nuclear cross section is $\sigma_N=
\sigma_{el}^H+\sigma_{el}^A+\sigma_{in}^A$. The probability of a nuclear
reaction of a proton that travels distance $\mathrm{d}t$ is

\begin{equation}
\mathrm{d}P_N=\sigma_N\mathrm{d}t,
\end{equation}

\noindent where $\sigma_N$ is in cm$^{-1}$ and $\mathrm{d}t$ is in cm.
If a proton undergoes a nuclear interaction, its energy and direction
of travel change, or it is absorbed. We therefore remove such protons 
from the primary beam after each step by applying to fluence $\Phi(E)$ 
the attenuation factor $\exp{[-\sigma_N(E})\Delta t]$, where $\Delta t$ 
is the step length. 

\subsubsection{Elastic scattering of protons on hydrogen atoms 
($p\rightarrow H$)} 

\hfill\\ 

Nuclear reactions contribute much less to dose than do Coulomb 
interactions with atomic electrons. Therefore, we use a different,
 more sparse, grid 
$\zeta_1,\ldots,\zeta_n$ for nuclear reactions than 
$z_1,\ldots,z_n$ that we use for multiple Coulomb scattering.
How these grids are designed is discussed in Subsection II.D.3.

Angular distributions of scattered protons in the center of mass 
frame are shown in Fig. 2. They are taken from 
Arndt\cite{Ar98}. We use
the isotropic scattering approximation $f_0(\mu_{cm})=1/2$, where 
$\mu_{cm}$ is the cosine of scattering angle in the center of mass frame
and $f_0(\mu_{cm})$ is the probability density. The latter is shown with 
a red dashed line in Fig. 2. This is a sufficiently accurate 
approximation, because this process is relatively rare. If a more 
accurate model of this process is needed, a weight 
$w=f(\mu_{cm})/f_0(\mu_{cm})$ is assigned to the particle when it 
scatters by an angle $\mu_{cm}$, where $f(\mu_{cm})$ is the accurate
 angular distribution (Fig. 2).

Nuclear reactions contribute much less to dose than do Coulomb 
interactions. Therefore, we use a different, more sparse, grid 
$\zeta_1,\ldots,\zeta_n$ for nuclear reactions than 
$z_1,\ldots,z_n$ that we use for multiple Coulomb scattering.
How these grids are designed is discussed in Subsection 2.4.3.

Angular distributions of scattered protons in the center of mass 
frame are shown in Fig. 2. They are taken from Arndt (1998). We use
the isotropic scattering approximation $f_0(\mu_{cm})=1/2$, where 
$\mu_{cm}$ is the cosine of scattering angle in the center of mass frame
and $f_0(\mu_{cm})$ is the probability density. The latter is shown with 
a red dashed line in Fig. 2. This is a sufficiently accurate 
approximation, because this process is relatively rare. If a more 
accurate model of this process is needed, a weight 
$w=f(\mu_{cm})/f_0(\mu_{cm})$ is assigned to the particle when it 
scatters by an angle $\mu_{cm}$, where $f(\mu_{cm})$ is the accurate
 angular distribution (Fig. 2).

\begin{center}
 \includegraphics[width=110mm]{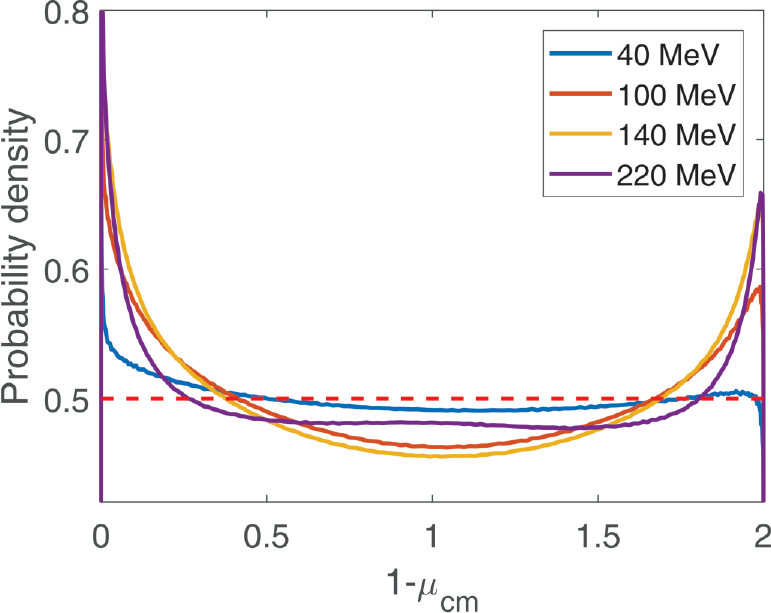}
 \end{center}
 \begin{flushleft}
    Figure 2. Angular distribution in the center of mass frame
   for elasic $p \rightarrow H$ interactions. Energies of the 
   incident proton before the collision are 40, 100, 140, and 220 MeV. 
   The red dashed line shows the isotropic scattering approximation, 
   $f_0(\mu_{cm})=1/2$.
 \end{flushleft}
 \vspace{0.2cm}

Elastic $p \rightarrow H$ scattering produces two protons, primary and
recoil. If the cosine of the scattering angle in the center of mass 
frame is $\mu_{cm}$, then the directional cosines of the two protons in 
the laboratory frame are

\begin{equation} 
\mu_{lab, 1}=\sqrt{\frac{1+\mu_{cm}}{2}},
\end{equation}

\begin{equation} 
\mu_{lab, 2}=\sqrt{\frac{1-\mu_{cm}}{2}},
\end{equation}

\noindent and their kinetic energies are

\begin{equation} 
\epsilon_i=\mu_{lab, i}^2 E, \:\:\:\:\:\: i=1,2,
\end{equation}

\noindent where $E$ is the kinetic energy of the proton before the
collision. If the incident proton before a collision travelled parallel 
to the $z$-axis in the positive direction, then after a collision 
$\mu_{lab, 1}>0$ and $\mu_{lab, 2}>0$. This means that elastic 
$p \rightarrow H$ collisions do not produce backscattered particles. 

To calculate contributions to fluence spectra at a depth $\zeta$ from  
primary and recoil protons, $\Phi_{el,i}^H(\zeta,E)$, $i=1,2$, we use a 
method similar to the Monte Carlo surface tally\cite{Va17}. 
First, we calculate distance $l_i$ in the direction defined by the 
directional cosine $\mu_{lab, i}$, from a given point in particle path, 
$t$, to the plane normal to the $z$-axis and located at depth $\zeta$:

\begin{equation} 
l_i=\frac{\zeta-t}{\mu_{i,lab}},  \:\:\:\: i=1,2. 
\end{equation}

If distance $l_i$ exceeds the CSDA range $R(\epsilon_i)$ of a proton 
with energy $\epsilon_i$, then the fluence contribution is zero. 
Otherwise, proton energy $\epsilon_i^*$ at depth $\zeta$ is calculated 
using Eq. (11), and a contribution $q_i=1/\vert \mu_{lab, i} \vert$ is 
added to the respective energy bin of fluence spectrum tally histogram 
at depth $\zeta$.

For clarity, let us consider interval $z_0<\zeta<z_1$, where $z_0$ is 
the proton starting point, where it had energy $E_{init}$. We assume 
that in this interval all protons have energy $E_{init}$ excluding those
that underwent a nuclear interaction. Then,

\begin{equation*}
\Phi_{el,i}^H(\zeta,E \vert E_{init})=\int_0^{\zeta}
\sigma_{el}^H(t)\mathrm{d}t
\int_{-1}^{1}
\frac{A_i(t,\mu_{cm})}
{\vert \mu_{lab,i}(\mu_{cm})\vert}
\delta(E-\epsilon_i^*(t,\mu_{cm}))
\frac{\mathrm{d}\mu_{cm}}{2}, \:\:\:\; 
\end{equation*}
\begin{equation}
i=1,2.
\end{equation}

\noindent Notations: $\Phi_{el,i}^H(\zeta,E \vert E_{init})$ is the 
fluence spectrum at depth $\zeta$ produced by each of the two scattered 
protons ($i=1,2$); $\sigma_{el}^H(t)$ is the total cross section for 
elastic $p \to H$ scattering at point $t$, 
$\sigma_{el}^H(t)=\sigma_{el}^H(E_{init})$ for $t<z_1$; 
$\epsilon_i^*(t,\mu_{cm})$ is the energy of a proton with initial energy 
$\epsilon_i(\mu_{cm})$ after it travels distance $l_i$, $\epsilon_i^*$ 
is calculated using Eq. (11); $A_i(t,\mu_{cm})$ accounts for attenuation
of fluence of scattered protons due to nuclear interactions, as they 
travel distance $l_i$; $A_i$ is set to zero, if $l_i<0$ or 
$l_i>R(\epsilon_i)$, otherwise it is calculated as 
$A_i=\exp{\left[-\sigma_N(\epsilon_i^*)\cdot l_i\right]}$; this formula
is a simplification we make to save computing time, the accurate formula
involves integration of $\sigma_N$ over distance $l_i$; the meaning of 
the $\delta-$function $\delta(E-\epsilon_i^*(t,\mu_{cm}))$ is as 
follows: in actual calculations the left hand side of Eq. (47) is an 
energy histogram, and the $\delta-$function indicates contribution to an
energy bin to which $\epsilon_i^*$ belongs.

The integral over $t$ is the sum of contributions to the fluence 
spectrum at depth $\zeta$ from $p \rightarrow H$ collisions at all 
depths $t<\zeta$. The integral over $\mu_{cm}$ accounts for angular
distribution of scattered protons. It is calculated using the Legendre 
quadrature.

In a more general case, when $\zeta>z_1$ the integral over $t$ is a line 
integral calculated by making steps $\Delta t$ as if, similarly to Monte 
Carlo, we follow a particle along its path. Except, in our algorithm the 
steps are not random, but optimized for best performance. In this case 
we also use Eq. (47) but now we need to integrate also over energy 
distribution of primary protons. For example, to find the contribution
to fluence at depth $\zeta$ from $p\rightarrow H$ interactions within a 
step from $\zeta_j$ to $\zeta_{j+1}$ ($z_1<\zeta_j<\zeta_{j+1}<\zeta$),
we modify Eq. (47) as follows:

\begin{equation*}
\Phi_{el,i}^H(\zeta,E)=\int_0^\infty \Phi(\zeta_j,E^\prime)
\mathrm{d}E^\prime
\int_{\zeta_j}^{\zeta_{j+1}}
\sigma_{el}^H(t)\mathrm{d}t \:\: \times
\end{equation*}
\begin{equation}
\int_{-1}^{1}
\frac{A_i(t,\mu_{cm})}
{\vert \mu_{lab,i}(\mu_{cm})\vert}
\delta(E-\epsilon_i^*(t,\mu_{cm}))
\frac{\mathrm{d}\mu_{cm}}{2}, \:\:\:\; 
i=1,2,
\end{equation}

\noindent where $\Phi(\zeta_j,E^\prime)$ is the fluence spectrum of 
primary protons at depth $\zeta_j$. In Eq. (47) we calculated quantities 
in the right hand side assuming that proton energy before the collision 
was $E_{init}$. In Eq. (48) in those calculations we use $E^\prime$ 
instead of $E_{init}$.

\subsubsection{Elastic scattering of protons on atoms heavier than 
hydrogen ($p\rightarrow A$)} 

\hfill\\ 

Elastic scattering on atoms heavier than hydrogen is overall similar to
$p \to H$ scattering. The main difference is that now we have one 
scattered proton instead of two. We do not consider displacement of a 
recoil atom $A$, and scattering is highly anisotropic. The differential 
cross section for targets with an atomic mass number $A<62$ is 
\cite{Ra72,Ts15}: 

\begin{equation}
\sigma_{el}^A(E,\mu_{cm})=A^{1.63}\exp{(-14.5s A^{0.66})}+
                          1.4 A^{0.33}\exp{(-10s)},
\end{equation}

\noindent where $s$ is the invariant momentum transfer in 
(GeV/c)$^2$. This is how $s$ is calculated:

\begin{equation}
p_{lab}=\sqrt{(m_1+E)^2-m_1^2}.
\end{equation}

\begin{equation}
E_{tot}=m_1+E+m_2.
\end{equation}

\begin{equation}
p_{cm}=\frac{p_{lab}m_2}
{\sqrt{ E_{tot}^2-p_{lab}^2}}.
\end{equation}

\begin{equation}
s=2 p_{cm}^2(1-\mu_{cm}).
\end{equation}

\noindent Notations: $p_{lab}$, $p_{cm}$ are proton momenta in the 
laboratory and center of mass frames, in GeV/c; $m_1$, $m_2$ are the 
rest energies of a proton and an atom $A$; $E$ is proton kinetic energy 
in the laboratory frame. The main part (i.e. small angles, 
$\mu_{cm}>0.9$) of angular distributions given by Eq. (49) is 
approximately an exponential function of $1-\mu_{cm}$. This property 
dictated our choice of the quadrature. 

The fluence of scattered protons after an 
elastic $p \to A$ scattering is calculated using a formula similar to 
Eq. (47). The only difference is that we now replace the factor 
$f_0(\mu_{cm})=1/2$ with the normalized differential cross section, 
$\sigma_{el}^A(E,\mu_{cm})/\sigma_{el}^A(E)$, where $\sigma_{el}^A(E)$
is the total cross section for the reaction:

\begin{equation}
\Phi_{el}^A(\zeta,E \vert E_{init})\!=\!\int_0^{\zeta}
\!\!\mathrm{d}t
\int_{-1}^{1}\!\!\sigma_{el}^A(t,\mu_{cm})
\frac{A(t,\mu_{cm})}
{\vert \mu_{lab}(\mu_{cm})\vert}
\delta(E-\epsilon^*(t,\mu_{cm}))
{\mathrm{d}\mu_{cm}}. \:\:\:\; 
\end{equation}

The integral over $t$ is a line integral calculated by making steps 
$\Delta t$. The integral over $\mu_{cm}$  is calculated using the 
Laguerre quadrature. To calculate $\mu_{lab}$ and the kinetic energy 
$\epsilon$ of a proton after scattering, for a given  $\mu_{cm}$, we
use relativistic kinematics as follows. The total energy and 
relativistic momenta of the incident proton in the laboratory and center 
of mass frames, $p_{lab}$ and $p_{cm}$, are given by Eqs. (50)-(52). The 
incident proton travels parallel to the $z$-axis, and 

\begin{equation}
\vec{\beta}\equiv\frac{\vec{v}}{c}=
\left(\beta_{x},\:\beta_{y},\: \beta_{z}\right)=
\left(0,\:0,\:p_{lab}/E_{tot} \right).
\end{equation}

\begin{equation}
\gamma\equiv\frac{1}{\sqrt{1-{\beta}^2}}=
\frac{E_{tot,lab}}{\sqrt{E_{tot,lab}^2-p_{lab}^2}}.
\end{equation}

\noindent The momentum after scattering is:

\begin{equation}
\vec{p}_{cm}^{\:\:\prime}=
\left(p_{cm,x}^{\prime},\:p_{cm,y}^{\prime},\: 
      p_{cm,z}^{\prime}\right)=
\left(p_{cm}\sqrt{1-\mu_{cm}^2},\: 0,\: p_{cm}\mu_{cm}\right).
\end{equation}

\noindent The total energy after scattering is:
\begin{equation}
E_{tot,cm}^{\:\prime}=\sqrt{m_1^2+(\vec{p}_{cm}^{\:\:\prime})^2}.
\end{equation}

\noindent
The total energy and momentum are transformed from the center of 
mass frame to the laboratory frame as follows\cite{Ge94}:

\begin{equation}
E_{tot,lab}^{\:\prime}=\gamma\left[E_{tot,cm}^{\:\prime}-
           (\vec{\beta}\cdot\vec{p}_{cm}^{\:\:\prime})\right].
\end{equation}

\begin{equation}
\vec{p}_{lab}^{\:\:\prime}=\vec{p}_{cm}^{\:\:\prime}+\gamma\vec{\beta}
              \left[\frac{\gamma}{\gamma+1}
              (\vec{\beta}\cdot\vec{p}_{cm}^{\:\:\prime})-
              E_{tot,cm}^{\:\prime}\right].
\end{equation}

\noindent Kinetic energy of the scattered proton is
\begin{equation}
\epsilon_{lab}^\prime=E_{tot,lab}^{\:\prime}-m_1.
\end{equation}

\noindent Finally, the directional cosine of the scattered proton is:
\begin{equation}
\mu_{lab}^{\:\prime}=\frac{p_{lab,z}^{\:\prime}}
          {\sqrt{(p_{lab,x}^{\:\prime})^2+
           (p_{lab,y}^{\:\prime})^2+
           (p_{lab,z}^{\:\prime})^2}}.
\end{equation}

\subsubsection{Inelastic scattering of protons on atoms heavier than 
hydrogen ($p\rightarrow A$)} 

\hfill\\ 

Cross sections for inelastic $p \to H$ interactions in tissue are zero 
for proton energies below 300 MeV. For heavier target atoms, after an 
inelastic $ p \to A$ collision, the nucleus may emit gamma radiation, 
neutrons, low energy protons and heavy recoils\cite{IC00}. We account 
for energy transport only by protons. Heavy recoils are assumed to 
deposit energy locally, and neutral particles escape the volume without
interactions. Total and single differential cross sections 
$\sigma_{in}^A(E)$ and $\sigma_{in}^A(E,\epsilon)$, respectively were 
taken from ICRU Report 63\cite{IC00}. Contributions to proton fluence 
from nonelastic interactions were calculated by a formula similar to 
Eq. (54):

\begin{equation*}
\Phi_{in}^A(\zeta,E \vert E_{init})\!=\!\int_0^{\zeta}
\!\!\mathrm{d}t \int_{0}^{E_{init}}
\!\!\sigma_{in}^A(E_{init},\epsilon)\mathrm{d}\epsilon \: \times
\end{equation*}
\begin{equation}
\int_{-1}^{1}\!\!\
\frac{A(t,\mu_{cm},\epsilon)}
{\vert \mu_{lab}(\mu_{cm})\vert}
\delta(E-\epsilon^*(t,\mu_{cm},\epsilon))
\frac{\mathrm{d}\mu_{cm}}{2} \: .
\end{equation}

In this process the scattering angle does not uniquely determine the
energy of the emitted proton $\epsilon$. For this reason Eq. (63) 
includes integration over $\epsilon$. In this equation we assumed 
isotropic angular distribution of emitted protons. A more accurate 
modelling of the angular distribution can be implemented using double 
differential cross sections that are also included in ICRU Report 63 
\cite{IC00}. After preliminary testing we concluded that a good balance 
of computing speed and accuracy is achieved without modeling this 
process. It was not included in the calculations presented in this 
paper.

\subsection{Algorithm implementation} 
\subsubsection{Calculations in heterogeneous media} 

\hfill\\

For treatment planning dose calculations the patient anatomy is
usually represented by a voxelized phantom. The voxel size is 2-4 mm, 
and the medium within a voxel is homogeneous. In our method the 
integration steps $\Delta t$ may span more than one voxel. Then for 
calculations in heterogeneous media a few modifications of the algorithm 
are needed. First, range calculations (Eq. (11)) will need to account 
for heterogeneity. Second, multiplications by the step size $\Delta t$, 
such as those in Eqs. (12), (26), (31) need to be replaced by 
integration over distance $\Delta t$. This integration is simple and 
fast, because in a voxelized phantom all the integrands are 
piecewise-constant functions. 

\subsubsection{\textit{Calculation of absorbed dose and average LET}} 

\hfill\\ 

Dose calculation is a postprocessing step. It is done after all the
fluence spectra have been calculated:

\begin{equation}
D(\vec{r}\:)=\frac{1}{\rho(\vec{r}\:)}\int_0^\infty
           \Phi(\vec{r},E)S(\vec{r},E)\mathrm{d}E, 
\end{equation}

\noindent where $\rho$ is the medium density and $S$ is the proton 
stopping power for the material at point $\vec{r}$. In this formulation, 
it is easy to calculate, if needed, dose to water instead of dose to the 
medium. In that case, the stopping power for water and water density are 
used in the above formula.

An advantage of our DBS is that it calculates fluence spectra that 
can be used for radiobiological modelling and for biological
optimization of treatments. Some radiobiological models, for example, 
require the frequency or dose average LET. With our method they are 
calculated as easily as the dose:

\begin{equation}
L_{F}(\vec{r}\:)={\int_0^\infty\Phi(\vec{r},E)L(\vec{r},E)}\mathrm{d}E/
             {\int_0^\infty\Phi(\vec{r},E)}\mathrm{d}E;
\end{equation}

\begin{equation}
L_{D}(\vec{r}\:)={\int_0^\infty\Phi(\vec{r},E)L^2(\vec{r},E)}\mathrm{d}E/
             {\int_0^\infty\Phi(\vec{r},E)L(\vec{r},E)}\mathrm{d}E,
\end{equation}

\noindent where $L(\vec{r},E)$ is proton LET for the material at point 
$\vec{r}$. Fluence spectra is a characteristic of proton beams 
sufficient for RBE calculations using also more advanced models based on
other than LET quantities, for example microdosimetric spectra.

\subsubsection{Spatial discretization} 

\hfill\\

Discretization of the spatial, angular and energy variables strongly
affects software performance. 

\hfill\\

\noindent {\it Multiple Coulomb scattering.} Step size 
$\Delta t_i=z_{i+1}-z_i$ for calculation of fluence spectra and angular 
distributions using iterative procedures given by Eqs. 2 and 17 is 
chosen so that a proton looses a fraction $f$ of its energy $E_i$ at
the step start, as it travels distance $\Delta t_i$. For a given $f$ the 
step size is calculated as follows:

\begin{equation}
\Delta t_i = R(E_i)-R((1-f) E_i),
\end{equation}

\noindent where $R(E_i)$ is the range of a proton with an initial energy 
$E_i$. The optimal value of $f$ is 0.05 for all proton energies that we 
have tested, 40-220 MeV. This method, however, produces very small steps 
near the track end. To correct this we introduced the minimal step size 
$\Delta t_{min}$. If in Eq. (67) $\Delta t_i$ becomes less than 
$\Delta t_{min}$, we set $\Delta t_i=\Delta t_{min}$. Hence, beyond a 
certain depth, all steps are the same, $\Delta t_{min}$. The optimal 
value of $\Delta t_{min}$ increases with increasing proton energy and is 
in the range of 0.005-0.2 cm.

{\it Nuclear interactions.} These processes are much less frequent than
Coulomb scattering. Hence, we use a more coarse grid. The step size
$\Delta\tau_i=\zeta_{i+1}-\zeta_i$ for nuclear interactions modelling 
(Eqs. (47), (48), (54), (63)) is set to $\Delta\tau_{max}$ at shallow 
depths, and then it is gradually reduced to $\Delta\tau_{min}$ towards 
the track end. Optimal values of both $\Delta\tau_{max}$ and 
$\Delta\tau_{min}$ increase with increasing proton energy. 
$\Delta\tau_{max}$ is in the range of 0.1-2.4 cm and $\Delta\tau_{min}$ 
is in the range of 0.1-0.8 cm.

\subsubsection{Energy discretization} 

\hfill\\

Proton fluence spectra for a monoenergetic source are very narrow at 
shallow depths and widen as protons slow down. For optimal 
performance we designed an energy scale that has a high resolution 
($\Delta E_{min}$) at high energies and gradually transitions to a lower 
resolution ($\Delta E_{max}$) for lower energies. The grid is generated 
as follows:

\begin{equation}
E_1=0;
\end{equation}

\begin{equation}
E_i=\Delta E_{max}\sum_{i=0}^{i-2}q^i, \: 2\le i \le N;
\end{equation}

\begin{equation}
E_i=E_N+\Delta E_{min}\cdot(i-N), \: i>N, \: E_i\le E_{init}.
\end{equation}

\noindent Notations: $q=(E_s-\Delta E_{max})/(E_s-\Delta E_{min})$;
$E_s$ is the energy at which the energy scale switches from variable 
steps (Eq. (69)) to a fixed step size (Eq. (70)), $E_s \approx E_N$,
$N=\lfloor\left\{2+\ln{(\Delta E_{min}}/
\Delta E_{min})/\ln{q}\right\}\rfloor$; $E_{init}$ is the initial proton 
energy. All three parameters that define the energy grid, 
$\Delta E_{max}$, $\Delta E_{min}$ and $E_s$ increase with increasing 
initial proton energy. $\Delta E_{max}$ is in the range  0.3-0.5 MeV, 
$\Delta E_{min}$ is in the range 0.02-0.1 MeV and $E_s$ is in the range 
31-175 MeV.

\subsubsection{Angular discretization} 
\hfill\\

We use a uniform on the logarithmic scale grid that starts at 
$\theta_{min}=0.01$\textdegree and spans to 
$\theta_{max}=10$\textdegree. The grid length is 50.

\subsubsection{Discretization of radial distance} 
\hfill\\

For the narrow Gaussian beam with the width $\sigma$=0.5 cm we used a 
nonuniform grid that starts at $r_{min}=0$ and spans to $r_{max}=2$ cm.
For the high energy of 220 MeV we extended the grid to $r_{max}=3$ cm.
Nodes $r_i$ of the grid are found by solving

\begin{equation}
1-\Delta y  \cdot (i-1)=\exp{\left(-\frac{r_i^2}{2\sigma^2} \right) };
\:\:\:\:\:i=1,2, \ldots i_{max}. 
\end{equation}

\noindent Here, $i_{max}$ is is the highest $i$ for which the left hand 
side is positive, $\Delta y$=0.025. Beyond $r_{imax}$ and up to 
$r_{max}$ the grid is uniform with the step 
$\Delta r=r_{imax}-r_{imax-1}$.

\section{Results and Discussion} 

The software is written in Fortran 95. We performed all the 
calculations on an HP Workstation with an Intel Core i7-10700 CPU, 2.9 
GHz. For comparison, we repeated the same calculations using Monte Carlo 
software Geant4 with the physics list QGSP\_BIC, optimal for hadron 
therapy\cite{Ge18}. The medium in all the calculations 
is liquid water. Protons are incident normally on water surface.

\subsection{Fluence spectra}

We report results of fluence spectra calculations at various depths for 
a point monoenergetic monodirectional proton source ($\delta$-source). 
Energies of incident protons were 40 MeV (Fig. 3), 100 MeV (Fig. 4), 
160 MeV (Fig. 5) and 220 MeV (Fig. 6). The CPU time for  our DBS was
5-11 ms. CPU times for the Monte Carlo simulations were tens of hours.
All the spectra are normalized per one incident proton. We made no other
 normalizations, scaling or adjustments of any kind. 

\begin{center}
 \includegraphics[width=110mm]{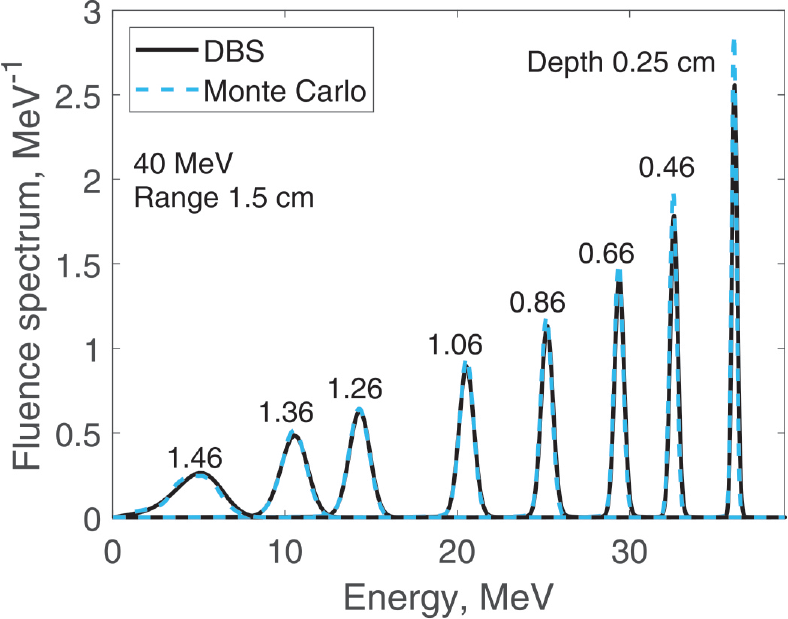}
 \end{center}
 \begin{flushleft}
    Figure 3. Fluence spectra for 40 MeV protons in water at several 
    depths as indicated in the figure. Comparison of our DBS with 
    Geant4 Monte Carlo results.
 \end{flushleft}
 \vspace{0.2cm}

\begin{center}
 \includegraphics[width=110mm]{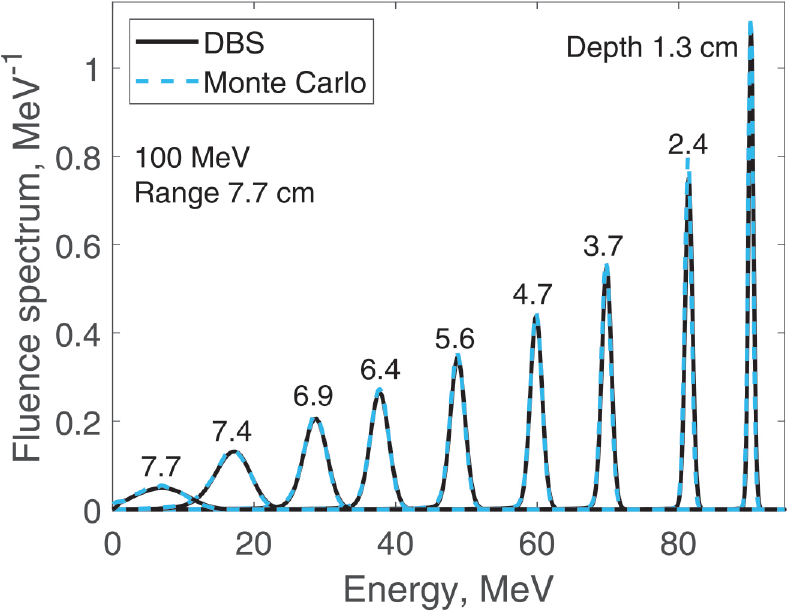}
 \end{center}
 \begin{flushleft}
    Figure 4. Fluence spectra for 100 MeV protons in water at several 
    depths as indicated in the figure. Comparison of our DBS with 
    Geant4 Monte Carlo results.
 \end{flushleft}
 \vspace{0.2cm}

\begin{center}
 \includegraphics[width=110mm]{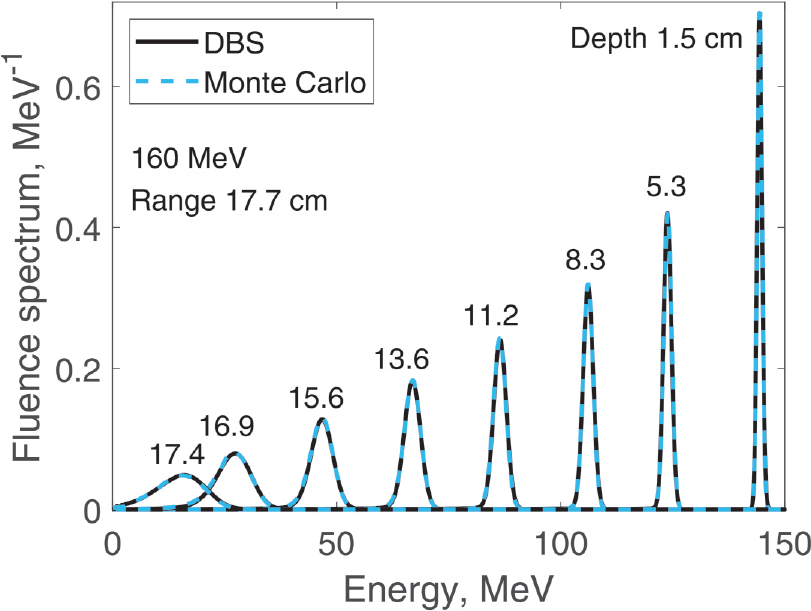}
 \end{center}
 \begin{flushleft}
    Figure 5. Fluence spectra for 160 MeV protons in water at several 
    depths as indicated in the figure. Comparison of our DBS with 
    Geant4 Monte Carlo results.
 \end{flushleft}
 \vspace{0.2cm}

\begin{center}
 \includegraphics[width=110mm]{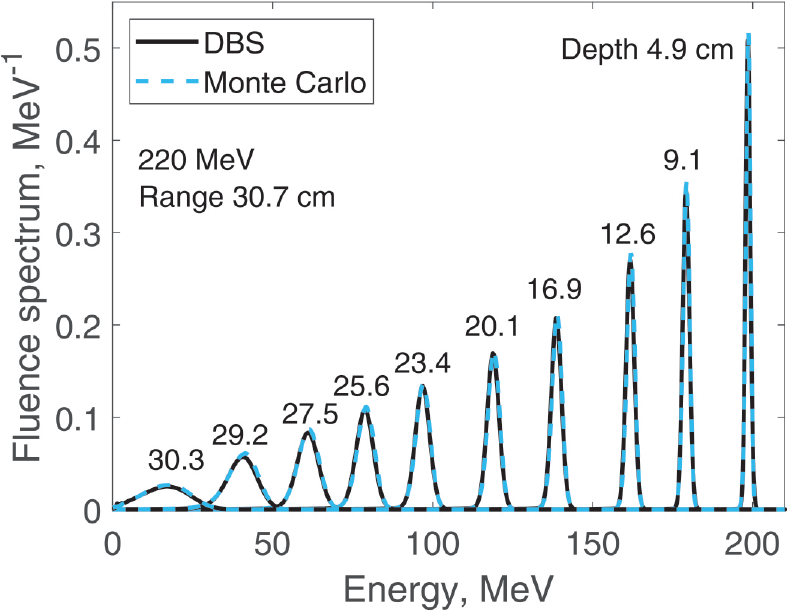}
 \end{center}
 \begin{flushleft}
    Figure 6. Fluence spectra for 220 MeV protons in water at several 
    depths as indicated in the figure. Comparison of our DBS with 
    Geant4 Monte Carlo results.
 \end{flushleft}
 \vspace{0.2cm}


\subsection{Depth dose} 

Dose distributions were calculated for the same four energies as the
spectra. In all the calculations, the incident proton fluence was the 
same for our DBS as it was for Monte Carlo. It was chosen so that in 
Monte Carlo simulations the entrance dose was 2 Gy. We made no other 
normalizations, scaling or adjustments of any kind. Depth doses are 
calculated at the postprocessing step. The CPU time was 0.1-1.6 ms. The 
calculation results are shown in Figs. 7 and 8.

\begin{center}
 \includegraphics[width=110mm]{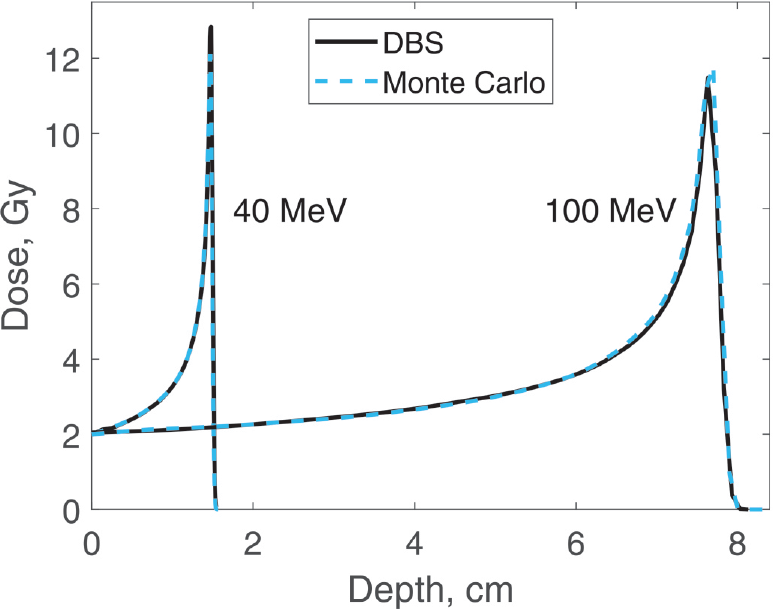}
 \end{center}
 \begin{flushleft}
    Figure 7. Dose versus depth for 40 MeV and 100 MeV protons in water. 
    Comparison of our DBS with Geant4 Monte Carlo results.
 \end{flushleft}
 \vspace{0.2cm}

\begin{center}
 \includegraphics[width=110mm]{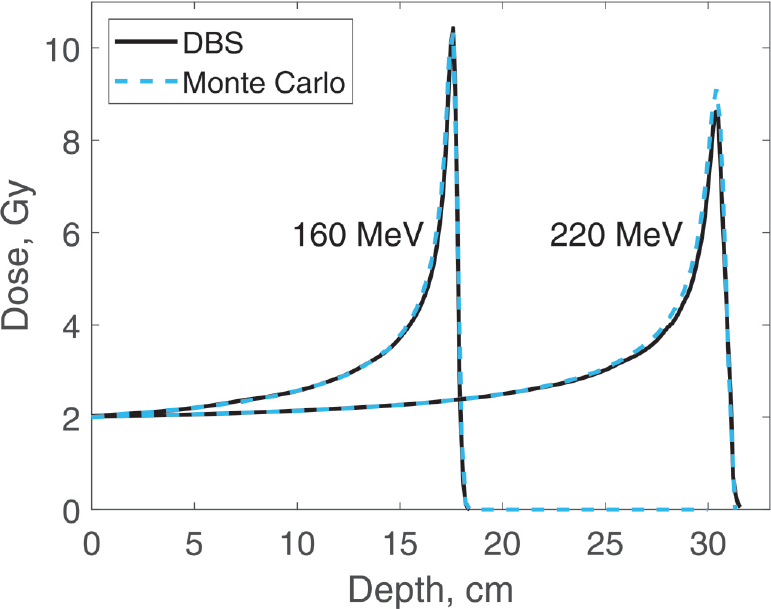}
 \end{center}
 \begin{flushleft}
    Figure 8. Dose versus depth for 160 MeV and 220 MeV protons in 
    water. Comparison of our DBS with Geant4 Monte Carlo results.
 \end{flushleft}
 \vspace{0.2cm}

\subsection{Narrow Gaussian beam. Dose distributions} 

The spatial distribution of the incident proton fluence is a 
two-dimensional Gaussian with $\sigma$=0.5 cm. Here we report on 
three-dimensional dose distributions for this beam. The incident proton 
fluence was the same for our DBS as it was for Monte Carlo. It was 
normalized so that for Monte Carlo simulations the entrance dose on the
 central axis was 1 Gy. We performed calculations for the same  four 
energies as in previous examples. In Figs. 9 and 10, for brevity, we 
show results only for 40 MeV and 220 MeV protons. In Table 1 we compare 
our DBS with Monte Carlo for all four energies using the $\gamma$-index 
test. The test included all grid nodes where the dose exceeded $0.01 
D_{max}$. The dose difference in this test is the difference between 
Monte Caro and DBS doses at the same point. The Table also shows the CPU 
time.

 \begin{center}
 \includegraphics[width=110mm]{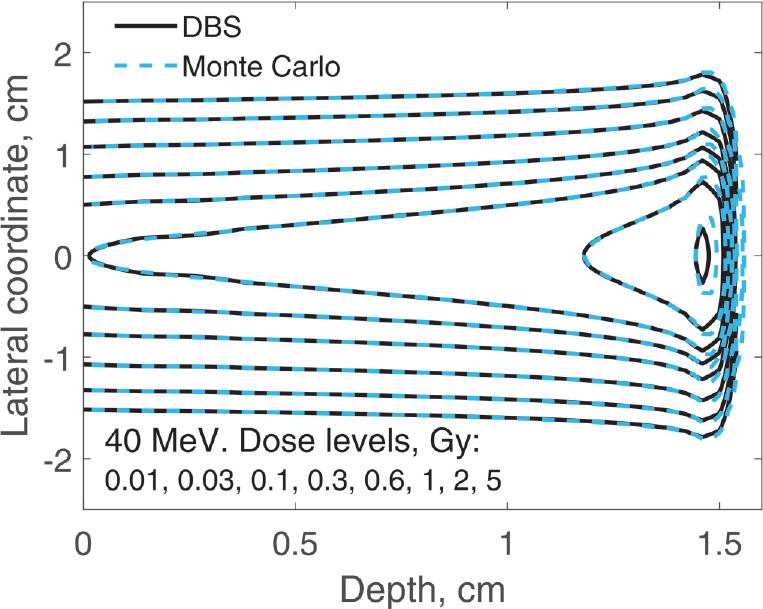}
 \end{center}
 \begin{flushleft}
    Figure 9. Dose distribution for a Gaussian beam with 
    $\sigma=$0.5 cm. The incident proton energy was 40 MeV.
    The isodose levels are given in the figure legend.
 \end{flushleft}
 \vspace{0.2cm}

 \begin{center}
 \includegraphics[width=110mm]{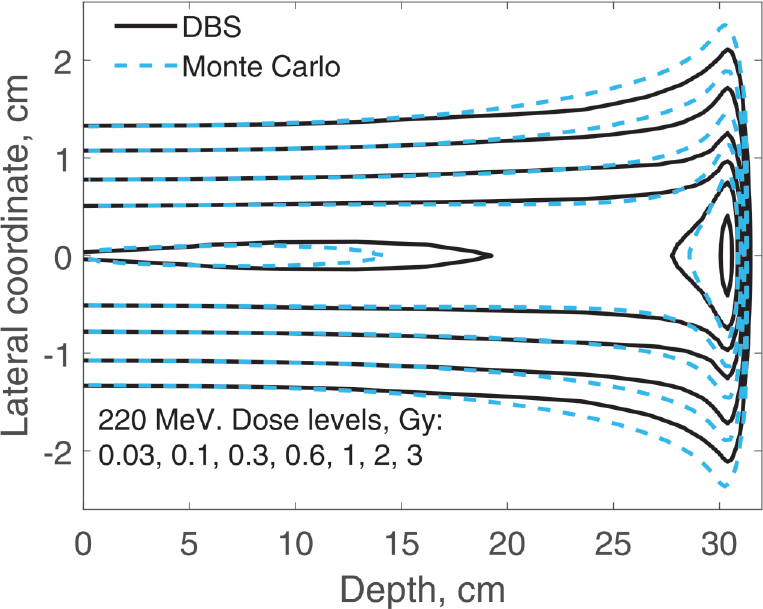}
 \end{center}
 \begin{flushleft}
    Figure 10. Dose distribution for a Gaussian beam with 
    $\sigma=$0.5 cm. The incident proton energy was 220 MeV.
    The isodose levels are given in the figure legend.
 \end{flushleft}
 \vspace{0.2cm}

   \begin{center}
   {Table 1. Comparisons of dose distributions, Figs. 9-10.}
   \vspace{0.1 cm}
   \end{center}
        \centering{     
       \begin{tabular}{|cllcc|}
\hline 
       & \multicolumn{2}{c}{$\gamma$-index test, fail rate} & & \\
E,     & \multicolumn{2}{c}{ volume fraction }   & $0.01D_{max}$, & CPU \\
\cline{2-3}\rule{-4pt}{15pt}
MeV    & 1\%/1 mm & 2\%/2 mm & cGy & time, ms\\
\cline{1-5}\rule{-4pt}{15pt} 
 40 & 0.0052 & 0  & 6.6 & 47 \\
100 & 0.010  & 0  & 5.4 & 31 \\
160 & 0.048  & 0  & 3.4 & 78 \\
220 & 0.050  & 0  & 1.9 & 34 \\
\hline               

\end{tabular}
}


\begin{justify}

\section{Conclusions}

We have developed and completed testing of a deterministic Boltzmann
equation solver (DBS) for dose and fluence spectra calculations for 
treatment planning of proton beam therapy. The DBS employs several 
innovative methods. It agrees mostly within 1\%/1 mm with one of the
most advanced Monte Carlo codes Geant4 with a physics list optimal 
for hadron therapy. We completed all our calculations in 5-78 ms on a
workstation with a modest CPU. 
Given the high computing speed of our DBS, and the generality of our 
approach, our DBS can be extended to include additional processes and 
implement alternative physical models to further improve the overall 
accuracy, if needed, or optimize performance for a particular type of 
problems. Also, our methods can be extended to heavier ions, such as 
carbon and helium. In contrast to other methods, our DBS provides 
accurate fluence spectra, for each node of a user defined spatial grid,
 thereby facilitating implementation of advanced RBE models that go 
beyond the basic quantities such as the average LET. This will help 
improve RBE models and advance the field of biological optimization of
treatment plans, which is particularly important for heavy ions. To 
summarize, our novel Boltzmann equation solver provides a foundation for
the development of fast and highly accurate treatment planning software
 for hadron therapy with protons and heavier ions.    

\section{Acknowledgements}
This study was supported by National Cancer Institute Grant R01 
CA225961. We thank Joe Munch of MD Anderson’s Research Medical Library 
for editorial assistance. 

\section{Competing Interests statement} 
Oleg Vassiliev and Radhe Mohan are inventors on PCT patent application
number PCT/US/2024/054392 related to this work filed by the University 
of Texas MD Anderson Cancer Center. The authors declare no other 
competing interests.

\end{justify}
\clearpage

\section*{References}
\addcontentsline{toc}{section}{\numberline{}References}
\vspace*{-20mm}





\bibliography{./dbs} 




\bibliographystyle{./medphy.bst}    


\end{document}